\documentclass[floatfix,showpacs,aps,pra, superscriptaddress, twocolumn]{revtex4-1}
\usepackage{hyperref}
\hypersetup{
	colorlinks=true,
	citecolor=blue,
	linkcolor=blue,
	urlcolor=blue}
\usepackage{graphicx}
\usepackage[normalem]{ulem}
\usepackage{amsmath}
\usepackage{amsfonts}
\usepackage{csquotes}
\usepackage{amssymb}
\usepackage{epsfig}
\usepackage{mathtools}
\usepackage{braket}
\usepackage[usenames,dvipsnames]{color}
\usepackage{setspace}
\usepackage{bm}
\usepackage{times}
\usepackage{ulem}
\usepackage{dsfont}
\usepackage{xcolor}

\begin{document}

\title{Excitations of a supersolid annular stripe phase in a spin-orbital-angular-momentum-coupled spin-1 Bose-Einstein condensate}
\author{Paramjeet Banger}\email{paramjeet.banger@acads.iiserpune.ac.in}
\affiliation{Department of Physics, Indian Institute of Technology Ropar, Rupnagar-140001, Punjab, India}
\affiliation{Department of Physics, Indian Institute of Science Education and Research Pune, Pune 411008, India}
\author{Rajat}\email{rajat.19phz0009@iitrpr.ac.in}
\affiliation{Department of Physics, Indian Institute of Technology Ropar, Rupnagar-140001, Punjab, India}
\author{Sandeep Gautam}\email{sandeep@iitrpr.ac.in}
\affiliation{Department of Physics, Indian Institute of Technology Ropar, Rupnagar-140001, Punjab, India}
\begin{abstract}
We present a theoretical study of the collective excitations of the supersolid annular stripe phase of a 
spin-orbital-angular-momentum-coupled (SOAM-coupled) spin-1 Bose-Einstein condensate. The annular 
stripe phase simultaneously breaks two continuous symmetries, namely rotational and $U(1)$ gauge
symmetry, and is more probable in the condensates with a larger orbital angular momentum
transfer imparted by a pair of Laguerre-Gaussian beams than what has been considered in the recent experiments. 
Accordingly, we consider a SOAM-coupled spin-1 condensate with a $4\hbar$ orbital angular momentum 
transferred by the lasers. Depending on the values of the Raman coupling strength and quadratic Zeeman term, the
condensate with realistic antiferromagnetic interactions supports three ground-state phases: the annular stripe, 
the vortex necklace, and the zero angular momentum phase. We numerically calculate the collective excitations of the condensate
as a function of coupling and quadratic Zeeman field strengths for a fixed ratio of spin-dependent and spin-independent
interaction strengths. At low Raman coupling strengths, we observe a direct transition from the zero angular momentum to the
annular stripe phase, characterized by the softening of a double symmetric roton mode, which serves as a 
precursor to supersolidity. 
\end{abstract}
\maketitle

\section{Introduction}
Unlike solids, where the electric field is crucial for generating spin-orbit 
(SO) coupling, ultracold neutral atoms do not exhibit SO coupling when exposed to an external electric field. 
Artificial gauge fields have emerged as a powerful tool which addresses this limitation, allowing the
study of the Lorentz-like force on neutral atoms~\cite{lin2009synthetic, lin2011synthetic}. The experimental realization of artificial gauge 
fields and the SO coupling that links spin with the linear 
momentum of neutral bosons~\cite{lin2009synthetic, lin2011synthetic, lin2011spin} have opened up new unexplored avenues of research in this field~\cite{galitski2013spin, Goldman_2014}. 
Synthetic SO coupling generated through Raman transitions has been successfully implemented in experimental setups for both bosonic \cite{lin2011spin,zhang2012collective, Campbell2016,luo2016}
and fermionic \cite{PhysRevLett.109.095302, PhysRevLett.109.095301, PhysRevLett.111.095301} atoms.
This coupling in spinor Bose-Einstein condensates (BECs) gives rise to various ground-state phases, such as 
the supersolid stripe, the zero momentum, and the plane wave phases, which have been extensively investigated~\cite{lin2011spin, Campbell2016,luo2016, 
li2017stripe, PhysRevLett.124.053605, wang2010spin, PhysRevLett.107.150403, PhysRevLett.110.235302, martone2016tricriticalities}. 
In the context of these ground-state phases, elementary excitations can provide fundamental insights, 
particularly through softening of the roton mode as a precursor to crystallization of the stripe phase~\cite{PhysRevA.90.063624,
PhysRevLett.114.105301, PhysRevA.93.033648, PhysRevA.93.023615}. Elementary excitations have also been utilized to map the ground-state
phase diagram in harmonically trapped SO-coupled BECs~\cite{zhang2012collective,  PhysRevA.95.033616, PhysRevLett.127.115301, *PhysRevLett.130.156001, PhysRevA.109.033319, rajat2024collective}.

Although SO coupling has been extensively studied in ultra-cold atoms, it is unlike the traditional SO coupling in atomic
physics, which refers to the interaction between spin and orbital angular momentum. 
In this context, the spin-orbital-angular-momentum (SOAM) coupling, which couples the atoms' spin and the orbital angular momentum, was theoretically
proposed~\cite{PhysRevA.91.033630, PhysRevA.91.053630, PhysRevA.92.033615, PhysRevA.91.063627, chen2016spin, vasic2016excitation, PhysRevA.96.011603}
and later experimentally realized \cite{PhysRevLett.122.110402,chen2018spin,chen2018rotating,liu2024vortexnucleationsspinorbose}.
In these experiments, co-propagating Gaussian and Laguerre-Gaussian (LG) laser beams along with a normally applied magnetic field have been used to
couple two \cite{PhysRevLett.122.110402} or three spin states of $^{87}$Rb \cite{chen2018spin,chen2018rotating} from the $F=1$ manifold 
to realize SOAM-coupled BECs and study their ground-state phases.

The primary ground-state phases of SOAM-coupled BECs can be broadly categorized as (a) the zero angular momentum (ZAM) phase, 
(b) the polarized phases, and (c) a variety of rotational symmetry-breaking phases 
\cite{PhysRevA.91.033630, PhysRevA.91.053630, PhysRevA.92.033615, PhysRevA.91.063627, chen2016spin, vasic2016excitation, duan2020symmetry, 
PhysRevResearch.2.033152, chiu2020visible, chen2020ground, bidasyuk2022fine, PhysRevA.108.043310, *bangerthesis}. The zero angular momentum (ZAM) and 
the polarized phases are axisymmetric
with well-defined angular momenta and have been observed experimentally \cite{PhysRevLett.122.110402,chen2018spin,chen2018rotating}. Among the
symmetry-breaking phases, an intriguing phase is the supersolid annular stripe (AS) phase, which corresponds to condensation in two
single-particle states with opposite angular momenta. This phase breaks two continuous symmetries, namely U(1) gauge and rotational
symmetry, just like the 
supersolid stripe phase breaks the gauge and translational symmetries in the Raman-induced SO-coupled BECs \cite{li2017stripe, PhysRevLett.124.053605}. Due to the small contrast and spatial period of stripes,
the AS phase has eluded the experimental detection. The effects of experimentally controllable parameters, such as angular momentum transferred 
to the atoms by the two LG beams, their waist sizes, the size of the BEC, and the interaction energy, on the feasibility of the experimental
detection of the AS phase have been theoretically studied \cite{chiu2020visible}.
It has been shown that larger angular momentum transfer by the LG beams can improve the spatial contrast of the stripes \cite{chiu2020visible}.
In this context, the experimental realizations of SOAM-coupling have been with an angular momentum transfer
of $\hbar$~\cite{PhysRevLett.122.110402,chen2018spin,chen2018rotating}, while ground-state phases of the SOAM-coupled BECs have been
examined with a larger momentum transferred by LG beams~\cite{ PhysRevA.91.063627, chiu2020visible, bidasyuk2022fine}.

In addition to the exploration of the equilibrium ground-state phases, elementary excitations 
in the zero angular momentum and the polarized phases of a pseudospin-1/2 SOAM-coupled spinor BEC have been theoretically 
investigated~\cite{PhysRevResearch.2.033152,vasic2016excitation,chen2020ground, PhysRevResearch.6.033200}. The low-lying modes in the excitation
spectrum, such as dipole and breathing modes, of the half-skyrmion and vortex-antivortex phases have been studied in
Refs.~\cite{PhysRevResearch.2.033152,vasic2016excitation}. For an SOAM-coupled spin-1 BEC with both antiferromagnetic
and ferromagnetic, the low-lying excitation spectrum of the polar-core vortex (ZAM) and coreless vortex (polarized) 
phases have been examined~\cite{PhysRevA.108.043310, *bangerthesis}. Regarding this study, it has to be emphasized that due to a smaller angular momentum 
transfer of $\hbar$ and the absence of the quadratic Zeeman field, the ground-state phase diagram did not have the AS phase.
The quadratic Zeeman field serves as an additional experimental controllable parameter in a spin-1 BEC, unlike in a pseudospin-1/2 BEC, and permits an alternative route to drive the phase transitions~\cite{chen2016spin, PhysRevA.93.033648, PhysRevA.93.023615, martone2016tricriticalities, rajat2024collective}.
In this work, we consider an SOAM-coupled spin-1 BEC with antiferromagnetic interactions and a higher angular momentum transfer of $4\hbar$ 
to allow for the emergence of the AS phase along with the ZAM and another symmetry-breaking vortex necklace (VN) phase~\cite{PhysRevResearch.5.023109, prykhodko2024vortexphasesdomainwalls}. 
We examine the collective excitations of the system as a function of quadratic Zeeman field and Raman coupling strengths and illustrate the signature of crystallization via a softening of a characteristic symmetric double roton mode at
the phase boundary.

The manuscript is organized as follows.
In Sec.~\ref{Sec-I}, we discuss the ground-state solutions and energy spectrum of 
the non-interacting SOAM-coupled Hamiltonian for a spin-1 BEC. In Sec.~\ref{Sec-III},
we present the interacting mean-field model and calculate the phase diagram in Raman coupling versus quadratic
Zeeman field strengths for an antiferromagnetic SOAM-coupled spin-1 BEC. 
In Sec.~\ref{Sec-IV}, we calculate collective excitations of the interacting SOAM-coupled spin-1 BEC and characterize a few 
low-lying modes. We provide a summary and conclusions of this study in Sec.~\ref{summary}.

\section{Non-interacting Hamiltonian}
\label{Sec-I}
We consider a non-interacting gas of SOAM-coupled spin-1 bosons in a quasi-2D harmonic trap.
The (dimensionless) single-particle Hamiltonian of the system in the plane polar coordinates is~\cite{chen2018spin,chen2018rotating,chen2016spin} 
\begin{align}
\label{single_pa}
H_0=&\left[-\frac{1}{2r}\frac{\partial}{\partial r}\left(r\frac{\partial}{\partial r}\right) + \frac{\hat{L}_z^2}{2}+\frac{r^2}{2}\right] 
+ \Omega(r) \left[\cos(l\phi) S_x\right. \nonumber\\&\left.-\sin(l\phi)S_y\right]  +q S_z^2,
\end{align}
where  $\hat{L}_z=-i \partial/\partial\phi$ is the angular momentum operator, $\Omega(r) = \Omega_0 {e}^{\frac{l}{2}}(r/w)^{l}e^{-2 r^2/w^2}$ is the
spatially dependent Raman coupling strength~\cite{chiu2020visible} with $\Omega_0$ and $w$ as the Rabi frequency and 
beam waists of the two LG beams, respectively, $q$ is the quadratic Zeeman term, and $S_x, S_y$ and $S_z$ are irreducible representations of the 
spin-1 angular momentum operators~\cite{chen2018spin,chen2018rotating}. Here, $l$ is the orbital angular momentum transfer to the atoms by the two co-propagating LG beams of opposite orbital angular momenta. We consider a higher orbital angular momentum transfer of $l=4$ with $w=5$ to allow for the emergence of the distinctive annular stripe (AS) phase in contrast to what was considered in Refs. \cite{chen2016spin,PhysRevA.108.043310, *bangerthesis}. By considering the unitary transformation of the Hamiltonian, with unitary 
operator $e^{-i lS_z\phi}$, the order parameter $\Psi = (\psi_{+1}, \psi_0, \psi_{-1})^T$ is transformed to 
$\Psi'=e^{-i lS_z\phi}{\Psi} =(e^{-il\phi}\psi_{+1}, \psi_0, e^{i l\phi} \psi_{-1})^T$, where $T$ denotes the transpose; the transformed Hamiltonian is
\begin{align}
\label{transformed_h}
{H^{'}_0} = & e^{-i lS_z\phi} H_0 e^{ i lS_z\phi},\nonumber\\ = & \left[-\frac{1}{2}\frac{\partial}{r \partial r}\left(r\frac{\partial}{\partial r}\right)+\frac{(\hat{L}_z+l S_z)^2}{2 r^2}+ \frac{r^2}{2}\right] \nonumber\\
&+\Omega(r)S_x+q S_z^2.
\end{align}
As $H^{'}_0$ commutes with $L_z$, a complete set of energy eigenstates can be chosen, which are the eigenstates of $L_z$ too. 
Accordingly, an arbitrary eigenstate of the single-particle Hamiltonian $H_0'$ can be defined as 
\begin{equation}\label{pop}
    {\Psi'(r,\phi)} = e^{il_z\phi}R(r),
\end{equation}
where $R(r) = [\psi_{+1}(r), \psi_{0}(r), \psi_{-1}(r)]^{T}$ is the radial part of the order parameter.
The orbital angular momentum $l_z = l_j \mp j l$ in the laboratory frame, where $l_j$ is the
angular momentum of the $j$th spin component \cite{chen2016spin,peng2022spin}. 

\begin{figure}[h]
\centering
\includegraphics[width=0.49\textwidth]{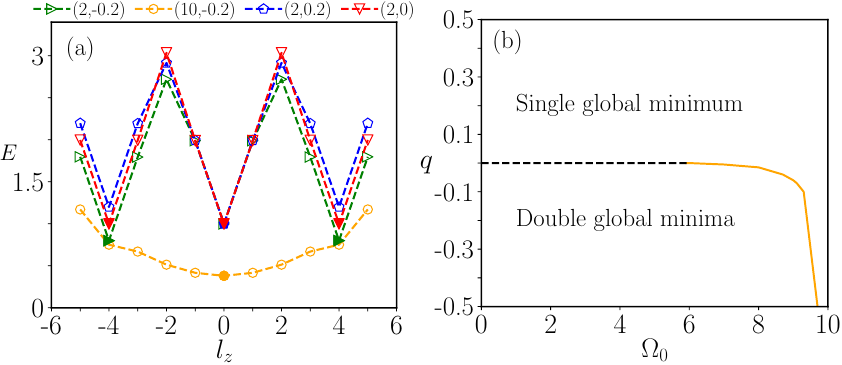}
\caption{(Color online) (a) The single-particle energy spectrum as a function of $l_z$ for $(\Omega_0, q) = (2, -0.2), (2,0), (2,+0.2)$ and $(10,-0.2)$. 
The spectrum has two degenerate global minima at $l_z=\pm4$ for $(\Omega_0,q)=(2,-0.2)$, three 
degenerate global minima at $l_z=\pm 4,0$ for $(\Omega_0,q)=(2,0)$, and a single global minimum at $l_z= 0$ for $(\Omega_0, q) = (2,+0.2)$ and $(10,-0.2)$.
(b) The single-particle phase diagram in the $\Omega_0,q$ plane. The single-particle spectrum is 
characterized by three degenerate global minima along the dashed part of the transition line.}
\label{sp_phase_diagram}
\end{figure}

\begin{figure}[h]
\centering
\includegraphics[width=0.46\textwidth]{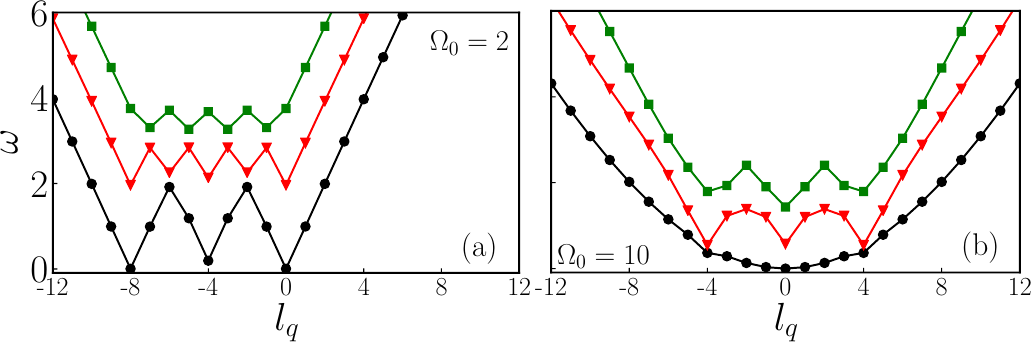}
\caption{(Color online) The single-particle excitation spectrum as a function of magnetic quantum number
$l_q$ associated with the angular momentum of the quasi-particle excitations: (a) $\Omega_0=2, q=-0.2$ and (b) $\Omega_0 = 10, q=-0.2$; 
the three lowest dispersion bands are shown. The lowest dispersion band has two global minima corresponding to two degenerate excitations with 
$l_q= -8$ and $l_q=0$ for $\Omega_0=2$, whereas it has single minimum located at $l_q= 0$ for $\Omega_0=10$. Refer to Sec.~\ref{Sec-IV}
for the methodology to calculate the excitation
spectrum.}
\label{single_particle_dispersion}
\end{figure}

We solve the single-particle Schrödinger equation $H_0' \Psi'(r,\phi)= E \Psi'(r, \phi)$ to calculate the energy
spectrum. The single-particle energy spectrum (equivalent to the lowest dispersion branch) for four pairs
of ($\Omega_0,q$) values is depicted in Fig.~\ref{sp_phase_diagram}(a). 
The spectrum is characterized by two global minima for
$\Omega = 2, q=-0.2$, a single global minimum for $\Omega = 2, q =0.2$ and $\Omega = 10, q =-0.2$. The single-particle
phase diagram shown in Fig.~\ref{sp_phase_diagram}(b) has single and two global-minima regimes separated by a 
phase boundary across a portion of which the spectrum has three degenerate minima \cite{chen2016spin}. 
We also solve the single-particle Bogoliubov de-Gennes (BdG) equation to determine the single-particle excitation
spectrum. For $q<q_c\le0$, where $q_c$ represents the critical quadratic Zeeman field above which the energy spectrum has a single
global minimum at $l_z = 0$, the lowest dispersion branch exhibits two degenerate global minima for smaller values of $\Omega_0$.
However, for larger values of $\Omega_0$, this branch has a single global minimum (see Fig.~\ref{single_particle_dispersion}) in
agreement with the single-particle energy spectrum.

\section{SOAM-coupled BEC with interactions}
\label{Sec-III}
At $T=0$ K, a weakly-interacting SOAM-coupled quasi-2D spin-1 BEC is very well described
by the Gross-Pitaevskii (GP) equation~\cite{chen2018spin,chen2018rotating,chen2016spin}  
\begin{equation}
i\frac{\partial \Psi}{\partial t} = \left(H_0 + \frac{c_0\rho}{2} + \frac{c_2 {\bf F}.{\bf S}}{2}\right)\Psi, \label{gpe}
\end{equation}
where
\begin{align} 
\rho = \Psi^{\dagger} \Psi = \sum_{j=\pm 1,0} |\psi_j(r,\phi)|^2, ~{\rm and}~{\bf F} = \Psi^\dagger{\bf S}\Psi\nonumber
\end{align}
with ${\bf S}$ denoting the vector of spin-1 matrices. In Eq.~(\ref{gpe}), $c_0$ and $c_2$ are interaction strengths
for the quasi-2D spin-1 BEC defined as
\begin{equation}
c_0 = \sqrt{8\pi\alpha}\frac{N(a_0+2a_2)}{3a_{\rm osc}},\quad c_2 = \sqrt{8\pi\alpha}\frac{N(a_2-a_0)}{3a_{\rm osc}},
\end{equation}
where $\alpha = \omega_z/\omega_r$ is the ratio of the axial to the radial frequency, $N$ is the total number of atoms in the condensate, $a_{\rm osc} = \sqrt{\hbar/{m\omega_r}}$, and $a_0$ and $a_2$ are the s-wave scattering lengths in total spin equal to 0 and 2 channels, respectively.
The interactions can lead to ground-state phases like the AS and the VN phases, which spontaneously break the rotational symmetry of the system.  
To realize the AS phase, in particular, we consider a $^{23}$Na BEC with antiferromagnetic interactions ($c_2<0$) in this work.  
We consider $10^5$ atoms of $^{23}$Na confined in an axisymmetric harmonic trap with $\omega_r = 2\pi \times 37$ Hz and $\omega_z = 2\pi \times 1000$ Hz, which tightly confines the system along the $z$-axis. The doublet of $s$-wave scattering lengths are 
$(a_0,a_2)=(50a_B,55.01a_B)$, where $a_B$ as the Bohr radius~\cite{crubellier1999simple}, 
and the corresponding interaction parameters are $c_0=42.57$, $c_2=1.33$.
We solve the time-independent version of the GP equation (\ref{gpe}) using imaginary-time propagation implemented via a 
time-splitting Fourier pseudospectral method~\cite{kaur2021fortress,*banger2022fortress,*banger2021semi,*ravisankar2021spin}.
The imaginary-time propagation, initiated with a suitable initial guess solution, facilitates a quick convergence towards the 
ground state. Inspired by the eigenfunctions of the single-particle Hamiltonian, we consider initial guess solutions of form 
$\Psi\sim e^{-(x^2+y^2)/2} \times [(x+iy)^{k+4},~-\sqrt{2}(x+iy)^{k},~(x+iy)^{k-4} ]^T$, where $k$ is an integer; additionally, 
we consider random guess solutions generated using a Gaussian random number generator.

\begin{figure*}[]
\centering
\includegraphics[width=0.8\textwidth]{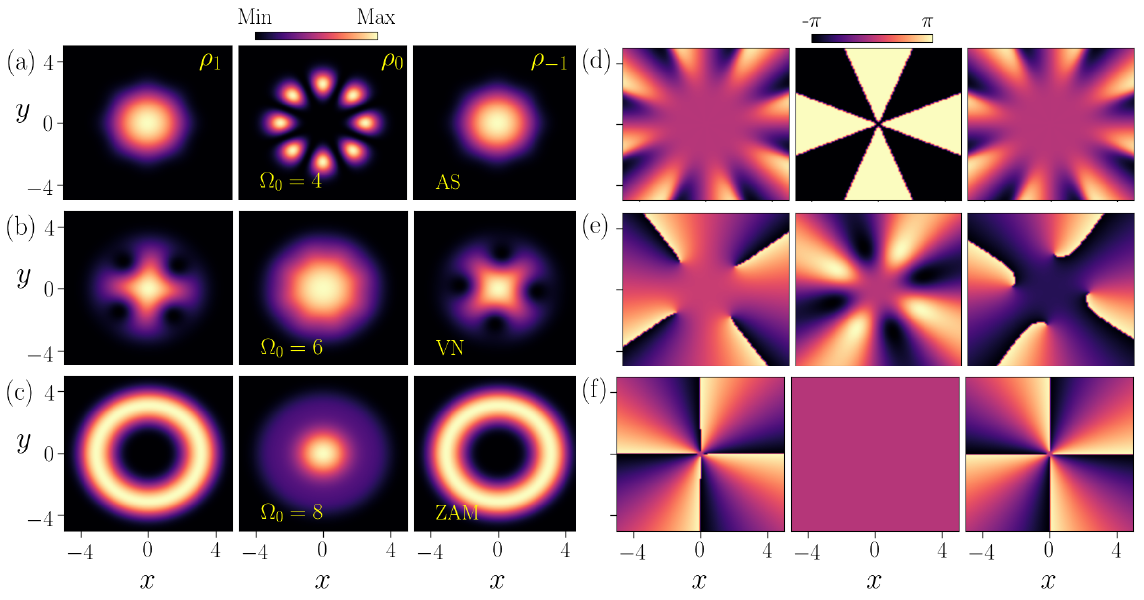}
\caption{(Color online) 
The density and the phase structures of the ground state phases of the SOAM-coupled spin-1 BEC with $c_0=42.57$, $c_2=1.33$ and $q=-0.2$.
The component densities in (a) the annular stripe (AS) phase with $ \Omega_0=4$, (b) the vortex necklace (VN) phase with $\Omega_0 = 6$, and (c) 
the zero angular-momentum (ZAM) phase with $ \Omega_0=8$. The phase profiles corresponding to (a), (b), and (c) are
displayed in (d), (e), and (f), respectively.}
\label{solution}
\end{figure*}
\begin{figure*}
\centering
\includegraphics[width=0.85\textwidth]{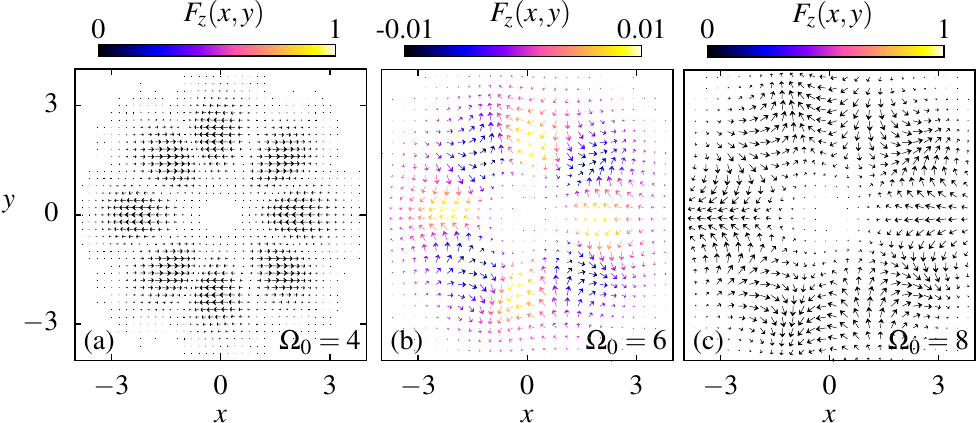}
\caption{(Color online) The spin-texture of SOAM-coupled spin-1 BEC with  $c_0=42.57$,  $c_2=1.33$ and $q=-0.2$: (a) $\Omega_0=4$, (b) $\Omega_0=6$, and (c) $\Omega_0=8$. The arrows show the projection of ${\bf F}(x,y)$ on the $x$-$y$ plane, and the color indicates its component along the $z$ axis.}
\label{spin_texture}
\end{figure*}

{\em Ground-state phases:} The system supports three ground-state phases in the $\Omega_0$-$q$ plane for $c_0 = 42.57$ and $c_2 = 1.33$, 
namely the AS, the VN, and the ZAM phase. The representative density and phase profiles of the phases are shown in Fig.~\ref{solution}. 
The AS and the VN phases break the rotational symmetry [see Figs. \ref{solution}(a) and (b)] {with $\int \Psi^\dagger(x,y) \hat L_z \Psi(x,y)dx dy=\langle \hat L_z\rangle \ne  0$}, whereas the ZAM phase is circularly symmetric with 
{$\langle\hat L_z \rangle = 0$ or zero-angular momentum per particle} [see Figs. \ref{solution}(c) and (f)]. The AS phase has a stripe pattern along the azimuthal direction in the density profile, and
the VN phase has four $\pm1$ charged phase singularities in $j = \pm 1$ component arranged along a circle. The ZAM phase is characterized by centrally located $4j$ phase singularity in the $j$th spin component. The ZAM phase, therefore, corresponds to the condensation occurring in a single particle state with $l_z = 0$, whereas the AS phase corresponds to the condensation in a superposition of two single-particle states with $l_z = +4$ and $-4$. The three phases have distinctive topological spin-texture ${\bf F}=(F_x, F_y, F_z)$ as shown in  Fig.~\ref{spin_texture}. 
The AS phase has $F_z =0$ with nonzero ${\bf F}\approx F_x\hat {\bf x}$ which is oppositely aligned in adjacent lobes of the texture and [see 
Fig.~\ref{spin_texture}(a)]. The VN and the ZAM phases have qualitatively similar projections of the spin textures on the $x$-$y$ plane, i.e., ${\bf 
F}_{\perp}$. However, $F_z$ is nonzero for the VN phase with the opposite signs in adjacent lobes of the texture and is zero for the ZAM phase and  
[see Figs.~\ref{spin_texture} (b) and (c)].
\begin{figure}[h]
\centering
\includegraphics[width=0.4\textwidth]{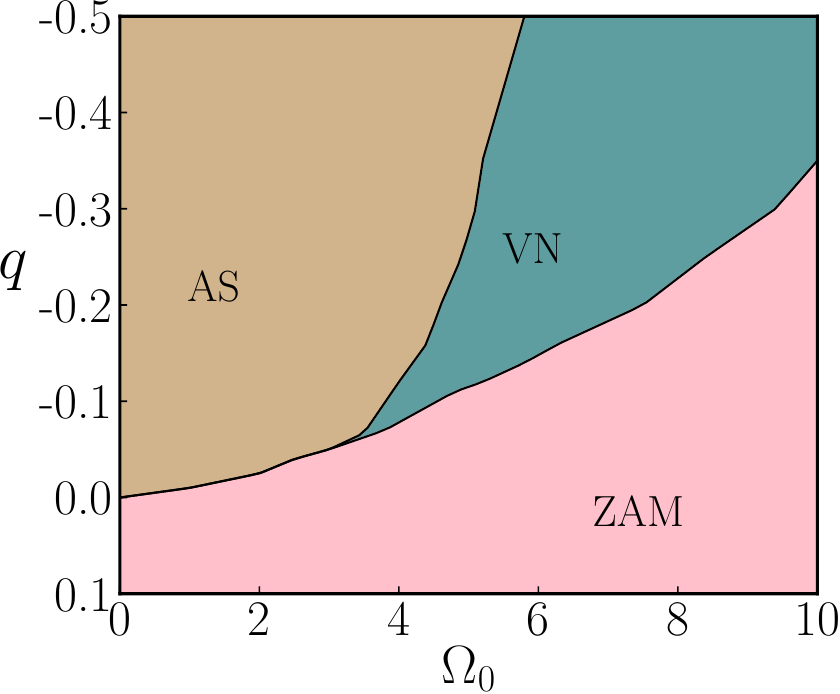}
\caption{(Color online) The phase diagram of an SOAM-coupled spin-1 BEC in the $\Omega_0-q$ plane with $c_0=42.57$
and $c_2=1.33$. The three phases coexist at the tricritical point 
$(\Omega_0 =3.5 , q =-0.06)$.} 
\label{phase_dia_ch5}
\end{figure}

The ground-state phase diagram in the $\Omega_0$-$q$ plane for the BEC with $c_0 = 42.57$ and $c_2=1.33$ is shown in Fig.~\ref{phase_dia_ch5}. 
For $q>0$, the ground state phase is the circularly symmetric ZAM phase with $l_z = 0$. For smaller Raman coupling strengths $\Omega < 3.5$, as
$q$ decreases (made more negative), there is a direct phase transition from the ZAM to the AS phase, whereas
for $\Omega > 3.5 $, the ZAM phase first transitions to the VN phase and then to the AS phase. The three phases coexist
at the tricritical point.
\begin{figure*}[!htpb]
\centering
\includegraphics[width=0.95\textwidth]{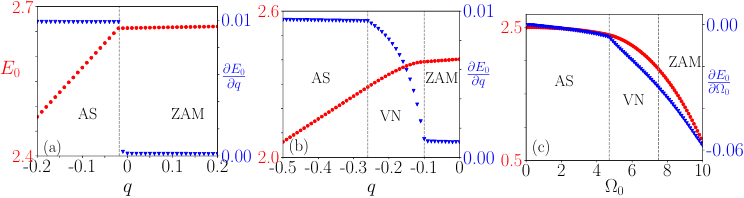}
\caption{The ground state energy $E_0$ and $\partial E_0/\partial q$ as a function of $q$ for the BEC with $c_0 = 42.57$, $c_2=1.33$: 
(a) $\Omega_0 = 2$ and (b) $\Omega_0 = 5$. (c) the ground state energy $E_0$ and $\partial E_0/\partial \Omega_0$ as a function of $\Omega_0$ 
for the BEC with the same interaction parameters. The dashed vertical lines correspond to the transition points. The first-order derivative of energy
changes discontinuously across the AS-ZAM transition point in (a), whereas it is continuous at the AS-VN and the VN-ZAM transition points in (b) and (c). 
}
\label{transition_order}
\end{figure*}
In the following section, we examine the collective excitations for (a) as a function of $q$ for $\Omega_0 = 2$, (b) as a 
function of $q$ for $\Omega_0 = 5$, and (c) as a function of $\Omega_0$ for $q=-0.2$. The variation of energy and its first-order derivative for these
three cases are shown in Figs.~\ref{transition_order}(a)-\ref{transition_order}(c). The direct phase transition from the
ZAM phase to the AS phase is a first-order phase transition
[see Fig.~\ref{transition_order}(a)], whereas, in the presence of the intervening VN phase, the AS-VN and the VN-ZAM phase transitions 
are continuous [see Figs.~\ref{transition_order}(b) and \ref{transition_order}(c)].

\section{Collective excitations}
\label{Sec-IV}
We use the Bogoliubov approach to calculate the collections excitation spectrum of the SOAM-coupled BEC, where we first linearize the  GP equation (\ref{gpe}) by perturbing the order parameter as
\begin{equation}\label{pop}
    {\Psi}(x,y,t) = e^{-i\mu t} [\Psi_{\rm eq}(x,y) +\delta {\Psi}(x,y,t)],
\end{equation}
where $\Psi_{\rm eq}(x,y)=[\psi_{+1}(x,y), \psi_{0}(x,y), \psi_{-1}(x,y)]^{T}$ 
is the ground-state order parameter, and $\mu$ is the chemical 
potential. We substitute the fluctuation $\delta {\Psi(x,y,t)}  = {\bf u}_{\lambda}(x,y) e^{-i\omega_{\lambda} t}
-{\bf v}_{\lambda}^*(x,y) e^{i\omega_{\lambda} t}$ 
in the linearized GP equation, where ${\bf u}_{\lambda}(x,y)$ and ${\bf v}_{\lambda}(x,y)$ are
Bogoliubov quasi-particle amplitudes and $\omega_{\lambda}$ is the excitation frequency with $\lambda$
as the frequency index. This leads to the following set of coupled Bogoliubov-de Gennes (BdG) 
equations
\begin{equation}
\begin{pmatrix}
P_1 & P_{2}\\
-P^*_2 & -P^*_1
\end{pmatrix} 
\begin{pmatrix}
{\mathbf u}_{\lambda} \\ {\mathbf v}_{\lambda} 
\end{pmatrix} 
=\omega_{\lambda}
\begin{pmatrix}
{\mathbf u}_{\lambda} \\{\mathbf v}_{\lambda} 
\end{pmatrix},
\label{bdg}
\end{equation}
where
$ 
{\mathbf u}_{\lambda} =(u_{+1,{\lambda}}, u_{0,{\lambda}}, u_{-1,{\lambda}})^{T},~ 
{\mathbf v}_{\lambda}=(v_{+1,{\lambda}}, v_{0,{\lambda}}, v_{-1,{\lambda}})^{T}\nonumber,
$
and $P_1$ and $P_2$ are defined in the Appendix.
We use a basis expansion method with a truncated set of eigenfunctions of a
two-dimensional harmonic oscillator serving as the requisite basis to solve the BdG equation
as discussed in the Appendix. 
The excitations can also be characterized by the magnetic quantum number $l_q$ for the circular-symmetric ZAM phase.
In this case, the GP equation can be linearized using the perturbed order parameter 
\begin{equation}
    {\Psi}(r,\phi,t) = e^{-i\mu t +i(l_z+S_z)\phi} [\Psi_{\rm eq}(r) +\delta {\Psi}(r,t)e^{i l_q \phi}], \nonumber\\
\end{equation}
which, followed by the Bogoliubov transformation, leads to a circularly symmetric set of BdG
equation~\cite{PhysRevA.108.043310, *bangerthesis}.  
 
\begin{figure}[!h]
\centering
\includegraphics[width=0.45\textwidth]{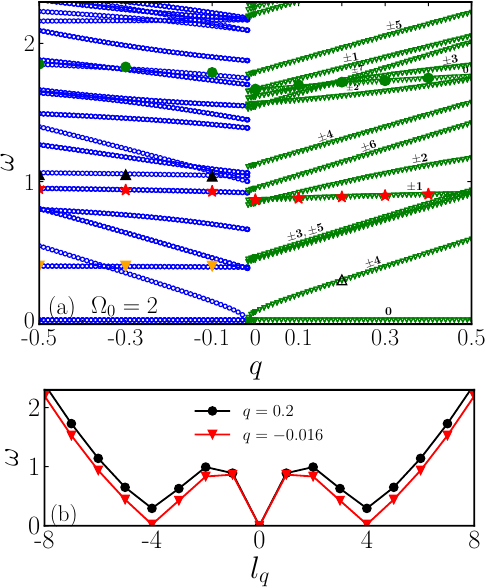}
\caption{(a) The low-lying excitation spectrum of the SOAM-coupled spin-1 BEC with  $c_0=42.57$, $c_2=1.33$,
and $\Omega_0=2$ as a function of $q$. For $q \gtrapprox -0.017$, there is a phase transition from the AS to the circularly
symmetric $l_z=0$ (ZAM) phases. In both the AS and the ZAM phases, dipole and breathing modes are marked by red stars
and green circles. Spin-dipole and spin-breathing modes are marked by orange lower and black upper triangles in the
AS phase. { For the circularly-symmetric ZAM phase, all the excitations are labelled with their respective 
magnetic quantum number $l_q$, which is not a good quantum number to characterize the excitations of the AS phase; the labels
have been placed above to their corresponding excitation frequencies.}
(b) Illustrates the excitation spectrum as a function of the magnetic quantum number $l_q$ in the ZAM phase,
where the roton minima occur at $l_q = \pm4$.}
\label{roton_mode}
\end{figure}
In Fig.~\ref{roton_mode}(a), we show the excitation spectrum of the BEC with $c_0 = 42.57 $ and $c_2 = 1.33$ as a function
of quadratic Zeeman field $q$ 
for Raman coupling strength $\Omega_0 = 2$. In this case, as $q$ is decreased, it leads to a direct phase transition from
the ZAM phase to the AS phase, as discussed in Sec.~\ref{Sec-III}. With a decrease in $q$, a double symmetric roton
mode~\cite{PhysRevA.93.033648, PhysRevA.93.023615, rajat2024collective} corresponding to
$l_q =\pm4$ softens and becomes zero at (critical) Zeeman field $q \approx -0.016$ as shown in  Fig.~\ref{roton_mode}(a). 
At this point, a transition occurs from the ZAM phase to the AS phase. This double symmetric roton structure is clearly visible in 
Fig.~\ref{roton_mode}(b), where we plot the excitation spectrum as a function of the magnetic quantum number $l_q$ 
for $q = 0.2$ and $q = -0.016$.
For the circularly symmetric ZAM phase, the modes
with $l_q\ne 0$ are doubly degenerate, whereas those with $l_q = 0$ are non-degenerate. This is a consequence of the invariance
of the BdG equation, under the transformation $l_q \rightarrow -l_q$ with a simultaneous interchange of $+1$ and $-1$ spin states 
for $l_z = 0$~\cite{PhysRevA.108.043310, *bangerthesis}.  
Some low-lying modes can be excited and identified by adding a suitable time-independent perturbation proportional to an observable 
$\hat O$ to the Hamiltonian at $t = 0$ and  then examining the time evolution ${\langle \hat O \rangle =} \int \Psi^\dagger(x,y,t) \hat O \Psi(x,y,t) d{\bf r}$,
where $\Psi(x,y,t=0)$ is the ground-state order parameter {of the unperturbed Hamiltonian}. The $\hat O$ can be chosen as $x$ or $y$ for the dipole,
$xS_z$ or $yS_z$ for the spin-dipole~\cite{PhysRevLett.127.115301}, $x^2$ or $y^2$ for the breathing and $x^2S_z$ or $y^2S_z$ for the spin-breathing 
mode. {
The $\langle \hat O \rangle$ oscillates at a dominant frequency, which can be extracted from its Fourier transform~\cite{PhysRevA.106.013304, PhysRevA.108.043310, *bangerthesis}. The density and spin dipole modes correspond, respectively, to the oscillations of the center of masses of 
$\rho(x,y)$ and $F_z(x,y)$, while the breathing modes refer to the oscillations of the root-mean-square 
sizes of these distributions.  
The method has been used experimentally to excite the density ~\cite{PhysRevLett.77.988} and spin dipole modes ~\cite{PhysRevA.94.063652} by modulating the 
harmonic trapping potential.}
The dipole and the breathing modes change discontinuously across the ZAM-AS phase boundary [see Fig.~\ref{roton_mode}(a)], 
highlighting the first-order nature of the transition in agreement with Fig.~\ref{transition_order}(a). The AS phase breaks two continuous symmetries, gauge and rotational symmetries, 
resulting in two zero-energy Goldstone modes in the excitation spectrum, whereas for the ZAM phase, we observe a single Goldstone mode due
to the breaking of $U(1)$ gauge symmetry. This first-order phase transition at low $\Omega$ is qualitatively similar to one studied
in an SO-coupled spin-1 BEC with antiferromagnetic interactions~\cite{rajat2024collective}. In both systems, the double symmetric roton mode, along with other low-lying modes, softens with a decrease in quadratic Zeeman field strength~\cite{rajat2024collective}, with vanishing roton gaps marking
the transition to the supersolid annular or rectilinear stripes.  

\begin{figure}[!h]
\centering
\includegraphics[width=0.45\textwidth]{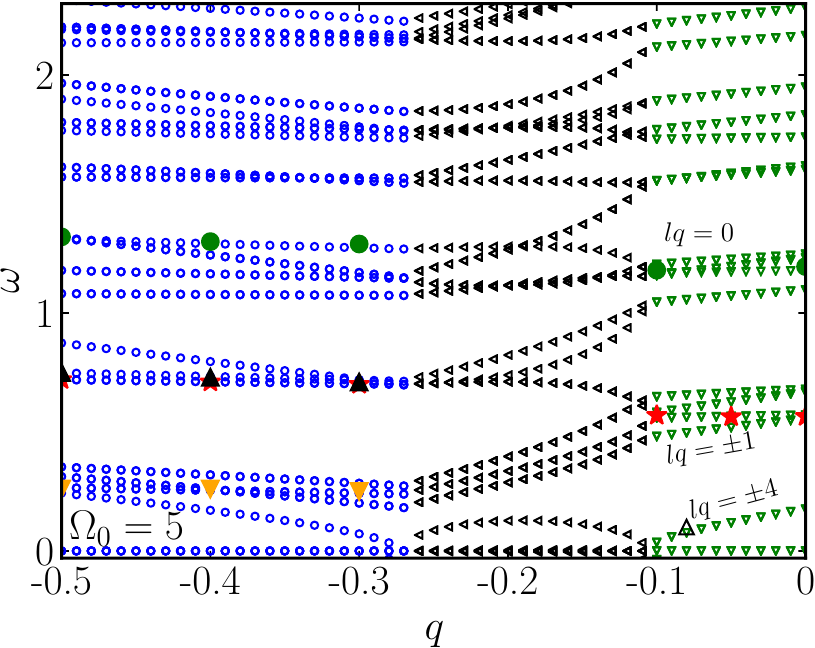}
\caption{(Color online) The excitation spectrum of the SOAM-coupled spin-1 BEC with  $c_0=42.57$, $c_2=1.33$, and $\Omega_0=5
$ as a function of $q$. The blue \enquote{circles}, black \enquote{triangles}
and green \enquote{downtriangles} correspond, respectively, to the annular stripe (AS), the vortex necklace (VN), and the zero angular momentum 
(ZAM) phases.
In the AS phase, dipole, breathing, spin-dipole, and spin-breathing modes are marked by
red stars, green circles, orange lower and black upper triangles, respectively.
In the ZAM phase, dipole ($l_q=\pm1$) and breathing modes ($l_q=0$) are marked by red stars and green circles.
For $q \gtrapprox -0.26$, there is a phase transition from the AS to the VN, and for $q \gtrapprox -0.1$, a phase transition from the VN to the ZAM phase. The excitation mode with $l_q=\pm4$ in the ZAM phase is the double roton mode. } 
\label{qmode}
\end{figure}
Next, we fix $\Omega_0$ at $5$ and examine the excitation spectrum as a function of $q$.
In this case, with a decrease in $q$ from 0 to -0.5, there is a phase transition from the ZAM to
the VN phase and then a transition from the VN to the AS phase [see Fig.~\ref{phase_dia_ch5}]
The excitation spectrum is shown in Fig.~\ref{qmode}.
Notably, both the transitions, from the AS to the VN and from the VN to the ZAM phase, 
are second-order transitions as evidenced by no discernible discontinuities across the critical points in Fig.~\ref{qmode}
and in agreement with the results in Fig.~\ref{transition_order}(b).
The dispersion ($\omega$ versus $l_q$) for the ZAM phase again has a symmetric double-roton structure (not shown here), with
roton gaps at $l_q=\pm 4$ closing at the ZAM-VN phase boundary.
Like the supersolid AS, the VN phase breaks the two continuous symmetries, resulting in two zero-energy Goldstone modes. In the VN phase, $\langle \hat O\rangle$ does not oscillate at a single 
dominant frequency for any of the $\hat O$ mentioned earlier, leading to the multiple peaks in the Fourier transform
of $\langle \hat O\rangle$. Due to this, we can not unambiguously identify dipole and breathing modes for
this phase.
\begin{figure}[!h]
\centering
\includegraphics[width=0.45\textwidth]{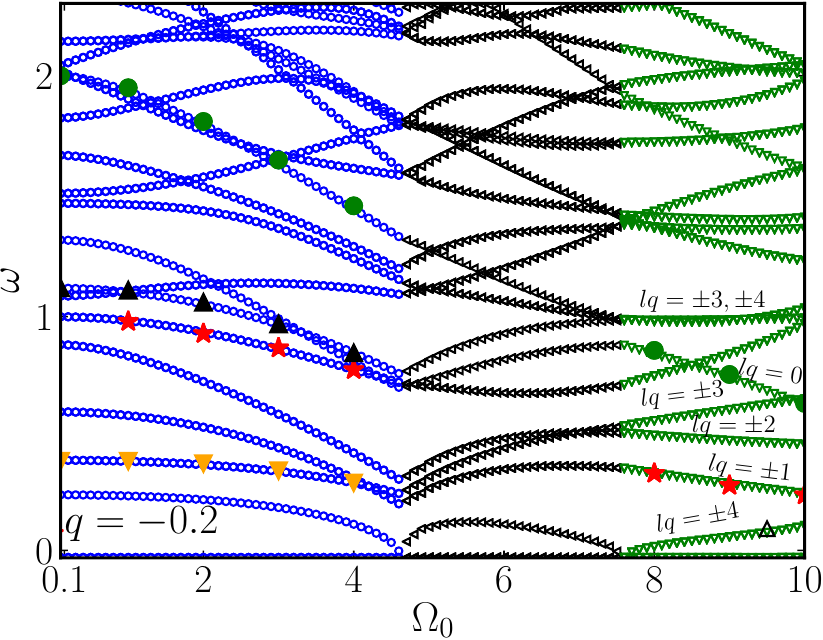}
\caption{(Color online) The excitation spectrum of the SOAM-coupled spin-1 BEC with  $c_0=42.57$, $c_2=1.33$, and $q=-0.2$ as
a function of $\Omega_0$. The blue \enquote{circles}, black \enquote{triangles} and green \enquote{downtriangles}
correspond, respectively, to annular stripe (AS), vortex necklace (VN), and zero angular momentum (ZAM) phases.
In the AS phase, dipole, breathing, spin-dipole, and spin-breathing modes are marked by
red stars, green circles, orange lower and black upper triangles, respectively.
In the ZAM phase, dipole ($l_q=\pm1$) and breathing ($l_q=0$) are marked by red stars and green circles.
For $\Omega_0 \gtrapprox 4.6$, there is a phase transition from the AS to the VN phase, and
at $\Omega_0 \gtrapprox 7.4$, a phase transition from VN to ZAM phases. The excitation mode with $\l_q =\pm4$ in the ZAM phase is the double roton mode and softens at the ZAM-VN transition point.}
\label{int_modes}
\end{figure}

Lastly, we keep the quadratic Zeeman field $q=-0.2$ fixed and vary the Raman coupling strength $\Omega_0$ 
from $0$ to $10$. As the Raman coupling strength increases, a second-order transition from the AS to the VN phase 
is observed above a critical coupling $\Omega_0\approx4.6$. As we further increase the Raman coupling strength, 
the system undergoes another continuous phase transition from the VN to the ZAM phase above $\Omega_0 \approx 7.4$. 
In the AS phase, the low-lying modes, namely the density dipole, density breathing, spin-dipole, and
spin-breathing modes, decrease with increasing $\Omega_0$. Similarly, in the ZAM phase, two density modes
decrease with increasing $\Omega_0$~[see Fig.~\ref{int_modes}]. 
 
\section{Summary and conclusions}
\label{summary}
We studied the ground-state phases and the low-lying collective excitations of a quasi-2D SOAM-coupled spin-1
BEC with antiferromagnetic interactions. We used the SOAM coupling corresponding to an angular
momentum transfer of $l=4\hbar$ to the atoms, which allows the annular stripe phase as one of the ground
state phases alongside a circularly symmetric (zero angular momentum) phase and another
symmetry-breaking vortex necklace phase for $c_2/c_0$ corresponding to $^{23}$Na. 
We calculated the phase diagram in the plane of Raman coupling
strength $\Omega_0$ versus the quadratic Zeeman field strength $q$ for both the non-interacting and interacting condensates. 
Using the Bogoliubov approach, we numerically calculated the excitation spectrum of the system with fixed interaction
strengths in three scenarios: (a) as a function of $q$ for a small fixed value of $\Omega_0$, (b) as
a function of $q$ for $\Omega_0$ fixed to a relatively higher value, and (c) as a function of
$\Omega_0$ for a fixed $q$. For (a), the excitation spectrum reveals a first-order phase transition
from the ZAM to the AS phase directly which is accompanied by the closing of double symmetric roton
gaps, a signature of crystallization or supersolidity. This is similar to the zero momentum to the
supersolid stripe phase transition in a SO-coupled BEC.
For (b) and (c), the continuous ZAM to the VN phase transition is also characterized by the closing double symmetric roton gaps. 
We identified a few low-lying collective modes, such as dipole and breathing modes, in both the AS and VN phases. 

\section*{ACKNOWLEDGMENTS}
S.G. acknowledges support from the Science and Engineering Research Board, Department
of Science and Technology, Government of India through Project No. CRG/2021/002597.

\section*{Appendix: A basis set expansion method to solve the BdG equations}
Here, we discuss the details of the numerical method to solve the BdG equation (\ref{bdg}), where
$P_1$ and $P_2$ are defined as
\begin{widetext}
\begin{align*}
{P}_{1}&=
\begin{bmatrix}
{\cal H} -\mu+c_{0} \rho_{+1}+{c_{2} (2 \rho_{+1}+\rho_{0}-\rho_{-1})} & (c_{0}+c_{2})\psi_0^*\psi_{+1}+2 c_2 
\psi_{-1}^*\psi_{0}+{\cal H}_{\Omega} & 
 {(c_{0}-c_{2})(\psi_{-1}\psi_{+1}^*)} \\
(c_{0}+c_{2})\psi_0\psi_{+1}^*+2 c_2 
\psi_{-1}\psi_{0}^*+{\cal H}_{\Omega}^* & {\cal H} -\mu +
c_{0} \rho_{0}+{c_{2}(\rho_{+1}+\rho_{-1})}&(c_{0}+c_{2})\psi_{0}\psi_{-1}^*+2c_2 
\psi_{+1}\psi_{0}^*+ {\cal H}_{\Omega}\\
 {(c_{0}-c_{2})(\psi_{+1}^*\psi_{-1})} &  
(c_{0}+c_{2})\psi_{0}^*\psi_{-1}+2c_2 
\psi_{+1}^*\psi_{0}+{\cal H}_{\Omega}^* & 
 {\cal H} -\mu+c_{0} \rho_{-1}+{c_{2}(\rho_{0}-\rho_{+1}+2 \rho_{-1})} \nonumber
\end{bmatrix},\\
{P}_{2}&=
\begin{bmatrix}
{(c_{0}+c_{2})\psi_{+1}^2} & {(c_{0}+c_{2})\psi_0\psi_{+1}} & 
{(c_{0}-c_{2})\psi_{-1}\psi_{+1}+c_{2}\psi_0^2} \\
{(c_{0}+c_{2})\psi_{+1}\psi_0} &
{c_{0}\psi_{0}^2+2 c_2 \psi_{+1} \psi_{-1}}&
{(c_{0}+c_{2})\psi_{-1}\psi_0} \\
{(c_{0}-c_{2})\psi_{+1}\psi_{-1}+c_{2}\psi_0^2} & {(c_{0}+c_{2})\psi_0\psi_{-1}} & 
(c_{0}+c_{2})\psi_{-1}^{2}
\end{bmatrix},
\end{align*}
\end{widetext}
with
\begin{align}
{\cal H} =&\left(-\frac{1}{2} \partial_x^2-\frac{1}{2} \partial_y^2+V(x,y)+c_{0} \rho \right),~\textrm{and}\nonumber\\
{\cal H}_{\Omega} =& \frac{\Omega(r)}{\sqrt{2}} e^{il\phi}\nonumber.
\end{align}
To solve the BdG equation (\ref{bdg}), we express BdG amplitudes $u_{j,\lambda}(x,y)$ and 
$v_{j,\lambda}(x,y)$ as a linear combination of $N_b$ low-lying eigenstates of 
two-dimensional harmonic oscillator \cite{roy-2020}
\begin{align}
u_{j,{\lambda}}(x,y) &= \sum_{n = 0}^{N_b-1} c_{j,n+1}^{\lambda} \varphi_n(x,y),\\
v_{j,{\lambda}}(x,y) &= \sum_{n = 0}^{N_b-1} d_{j,n+1}^{\lambda} \varphi_n(x,y),
\end{align}
where $j = +1,0,-1$ and $c_{j,n+1}^\lambda$ and $d_{j,n+1}^\lambda$ are the constant 
coefficients. The $n$th harmonic oscillator oscillator basis state is
\begin{equation}
\varphi_n(x,y) = \xi_{n_x}(x) \xi_{n_y}(y),
\end{equation}
where $\xi_{n_x}(x) [\xi_{n_y}(y)]$ is a normalized eigen state of one-dimensional harmonic 
oscillator with $n_x = 0,1,\ldots,n_x^{\rm max}$ $(n_y = 0,1,\ldots,n_y^{\rm max})$, and 
$n = n_y(n_x^{\rm max} + 1) + n_x$ for $n_y^{\rm max}\geq n_x^{\rm max}$. In this work, with an 
isotropic confinement along the $x-y$ plane, we consider $n_x^{\rm max} = n_y^{\rm max}$.
Projecting the six-coupled BdG equations on $N_b = (n_x^{\rm max}+1)^2$ harmonic oscillator 
states, we get $6N_b$ equations, which can be written in the $6\times6$ matrix form as
\begin{equation}\label{BdG_matrix_eqn}
\begin{bmatrix}
{\cal M}_{11} & {\cal M}_{12} & {\cal M}_{13} & {\cal M}_{14} & {\cal M}_{15} &  {\cal M}_{16} \\
{\cal M}_{21} & {\cal M}_{22} & {\cal M}_{23} & {\cal M}_{24} & {\cal M}_{25} &  {\cal M}_{26} \\
{\cal M}_{31} & {\cal M}_{32} & {\cal M}_{33} & {\cal M}_{34} & {\cal M}_{35} &  {\cal M}_{36} \\
{\cal M}_{41} & {\cal M}_{42} & {\cal M}_{43} & {\cal M}_{44} & {\cal M}_{45} &  {\cal M}_{46} \\
{\cal M}_{51} & {\cal M}_{52} & {\cal M}_{53} & {\cal M}_{54} & {\cal M}_{55} &  {\cal M}_{56} \\
{\cal M}_{61} & {\cal M}_{62} & {\cal M}_{63} & {\cal M}_{64} & {\cal M}_{65} &  {\cal M}_{66} 
\end{bmatrix}\begin{bmatrix}{\bf c}_{+1}^{\lambda}\\ {\bf c}_{0}^{\lambda}\\{\bf c}_{-1}^{\lambda}\\
{\bf d}_{+1}^{\lambda}\\ {\bf d}_{0}^{\lambda}\\{\bf d}_{-1}^{\lambda}\end{bmatrix} = \omega_{\lambda} 
\begin{bmatrix}{\bf c}_{+1}^{\lambda}\\ {\bf c}_{0}^{\lambda}\\{\bf c}_{-1}^{\lambda}\\
{\bf d}_{+1}^{\lambda}\\ {\bf d}_{0}^{\lambda}\\{\bf d}_{-1}^{\lambda}\end{bmatrix}.
\end{equation}
In Eq.~(\ref{BdG_matrix_eqn}), ${\bf c}_j^{\lambda}$ and ${\bf d}_j^{\lambda}$ are $N_b\times 1$
column vectors defined as
\begin{align} 
{\bf c}_j^{\lambda} &= (c_{j,1}^{\lambda}, c_{j,2}^{\lambda}, \ldots c_{j,N_b}^{\lambda})^T,\\
{\bf d}_j^{\lambda} &= (d_{j,1}^{\lambda}, d_{j,2}^{\lambda}, \ldots d_{j,N_b}^{\lambda})^T,
\end{align}
and the six elements of the block matrix $\cal M$ on the left hand side are $N_b\times N_b$ matrices
with their $kl$th element defined as follows
\begin{widetext}
\begin{eqnarray}
{\cal M}_{11}^{kl} &=& \iint \varphi_p(x,y)
\Bigg[-\frac{1}{2} \partial_x^2-\frac{1}{2} \partial_y^2-\mu +V(x,y)+c_{0} \rho +c_0 \rho_{+1} +{c_{2} (2 \rho_{+1} + \rho_{0}-\rho_{-1})} \Bigg]\varphi_q(x,y) dx dy,\nonumber\\
{\cal M}_{12}^{kl} &=& \iint \varphi_p(x,y)\Bigg[(c_{0}+c_{2})\psi_0^*\psi_{+1}+2 c_2 
\psi_{-1}^*\psi_{0}+h_{\rm cc}\Bigg]\varphi_q(x,y) dx dy,\nonumber\\
{\cal M}_{13}^{kl} &=& \iint \varphi_p(x,y)
{(c_{0}-c_{2})(\psi_{-1}\psi_{+1}^*)}\phi_q(x,y) dx dy,\:{\cal M}_{14}^{kl} = \iint \varphi_p(x,y)
{(c_{0}+c_{2})\psi_{+1}^2}\varphi_q(x,y) dx dy,\nonumber\\
{\cal M}_{15}^{kl} &=& \iint \varphi_p(x,y)
{(c_{0}+c_{2})\psi_0\psi_{+1}}\varphi_q(x,y) dx dy, \:{\cal M}_{16}^{kl} =\iint \varphi_p(x,y)\Bigg[ {(c_{0}-c_{2})\psi_{-1}\psi_{+1}+c_{2}\psi_0^2} \Bigg]\varphi_q(x,y) dx dy,\nonumber\\
{\cal M}_{21}^{kl} &=&\iint \varphi_p(x,y)
\Bigg[(c_{0}+c_{2})\psi_0\psi_{+1}^*+2 c_2 
\psi_{-1}\psi_{0}^*+h_{\rm cc}^*\Bigg]\varphi_q(x,y) dx dy,\nonumber\\
{\cal M}_{22}^{kl} &=& \iint \varphi_p(x,y)
\Bigg[-\frac{1}{2} \partial_x^2-\frac{1}{2} \partial_y^2-\mu +V(x,y)+c_{0} \rho +{c_{2} (\rho_{+1} + \rho_{-1})} \Bigg]\varphi_q(x,y) dx dy,\nonumber\\
{\cal M}_{23}^{kl}&=&\iint \varphi_p(x,y)\Bigg[
{(c_{0}+c_{2})\psi_{-1}^*\psi_{0}}+2c_2\psi_{0}^*\psi_{+1} + h_{\rm cc}\Bigg]\varphi_q(x,y) dx dy,\nonumber\\
{\cal M}_{24}^{kl}&=&\iint \varphi_p(x,y)
{(c_{0}+c_{2})(\psi_{+1}\psi_{0})}\varphi_q(x,y) dx dy,\nonumber\\
{\cal M}_{25}^{kl} &= &\iint \varphi_p(x,y)
\Bigg[c_{0}\psi_0^2+2 c_2 
\psi_{-1}\psi_{+1}\Bigg]\varphi_q(x,y) dx dy,\nonumber\\
{\cal M}_{26}^{kl} &= &\iint \varphi_p(x,y)
(c_{0}+c_2) 
\psi_{-1}\psi_{0}\varphi_q(x,y) dx dy,\nonumber\\
{\cal M}_{31}^{kl}& =& -\iint \varphi_p(x,y){(c_{0}-c_{2})\psi_{+1}^{*} \psi_{-1}}\varphi_q(x,y) dx dy,\nonumber\\
{\cal M}_{32}^{kl} &=&-\iint \varphi_p(x,y)
\Bigg[{(c_{0}+c_{2})\psi_0^*\psi_{-1}+ 2 c_{2} \psi_{+1}^* \psi_0 + h_{\rm cc}^*}\Bigg]\varphi_q(x,y) dx dy,\nonumber\\
{\cal M}_{33}^{kl} &=&- \iint \varphi_p(x,y)
\Bigg[-\frac{1}{2} \partial_x^2-\frac{1}{2} \partial_y^2-\mu +V(x,y)+ c_{0} \rho +c_{0} \rho_{-1} +{c_{2} (\rho_{0}-\rho_{+1}+2 \rho_{-1})} \Bigg]\varphi_q(x,y) dx dy,\nonumber\\
{\cal M}_{34}^{kl} &=& -\iint \varphi_p(x,y)
\Bigg[{(c_{0}-c_{2})\psi_{+1}\psi_{-1}}+c_2\psi_0^{2}\Bigg]\varphi_q(x,y) dx dy,\nonumber\\
{\cal M}_{35}^{kl} &=&- \iint \varphi_p(x,y)
\Bigg[(c_{0}+c_{2})\psi_{0}\psi_{-1}\Bigg]\varphi_q(x,y) dx dy,\nonumber\\
{\cal M}_{36}^{kl} &=& -\iint \varphi_p(x,y)
\Bigg[{(c_{0}+c_{2})\psi_{-1}^2}\Bigg] \varphi_q(x,y) dx dy,\nonumber\\
{\cal M}_{41}^{kl} &=& -({\cal M}_{14}^{kl})^*,\quad
{\cal M}_{42}^{kl}=-({\cal M}_{15}^{kl})^*,\quad
{\cal M}_{43}^{kl}=-({\cal M}_{16}^{kl})^*,\quad
{\cal M}_{44}^{kl} =-({\cal M}_{11}^{kl})^*,\quad
{\cal M}_{45}^{kl} =-({\cal M}_{12}^{kl})^*,\quad M_{46}^{kl}=-({\cal M}_{13}^{kl})^*,\nonumber\\
{\cal M}_{51}^{kl} &=& -({\cal M}_{24}^{kl})^*,\quad M_{52}^{kl} = -({\cal M}_{25}^{kl})^*,\quad
{\cal M}_{53}^{kl}=-({\cal M}_{26}^{kl})^*,\quad {\cal M}_{54}^{kl} = -({\cal M}_{21}^{kl})^*,\quad
{\cal M}_{55}^{kl} = -({\cal M}_{22}^{kl})^*,\quad
{\cal M}_{56}^{kl} = -({\cal M}_{23}^{kl})^*,\nonumber\\
{\cal M}_{61}^{kl} &=& -({\cal M}_{34}^{kl})^*,\quad
{\cal M}_{62}^{kl} = -({\cal M}_{35}^{kl})^*,\quad {\cal M}_{63}^{kl} = -({\cal M}_{36}^{kl})^*,\quad
{\cal M}_{64}^{kl}=-({\cal M}_{31}^{kl})^*,\quad {\cal M}_{65}^{kl} =-({\cal M}_{32}^{kl})^*,\quad
{\cal M}_{66}^{kl} = -({\cal M}_{33}^{kl})^*,\nonumber
 \end{eqnarray}
\end{widetext}
where $p$ and $q$ can have values $0,1,2,\ldots N_b-1$, $k=p+1$, and $l = q+1$.
We opt for a sparse matrix representation to store the BdG matrix and employ the ARPACK library \cite{doi:10.1137/1.9780898719628} for diagonalization.
LAPACK subroutines \cite{lapack} can also efficiently handle the diagonalization of the matrix for small $N_b$. 

\bibliography{bib_file}{}
\bibliographystyle{apsrev4-1}
\end{document}